\documentclass{article}
\usepackage{enumerate}
\usepackage{amsfonts}
\usepackage{amsmath}
\usepackage{amssymb}
\usepackage{amsthm}
\usepackage{latexsym}
\usepackage{color}

\def\A{\mathcal{A}}
\def\H{\mathcal{H}}

\def\S{\mathfrak{S}}
\def\C{\mathcal{S}}

\def\B{\mathfrak{B}}
\def\E{\mathcal{S}}
\newcommand{\supp}{\mathrm{supp}}
\newcommand{\rank}{\mathrm{rank}}
\newcommand{\id}{\mathrm{Id}}
\newcommand{\Tr}{\mathrm{Tr}}

\newcommand{\shs}{\hspace{1pt}}

\newcounter{defin}  \newcounter{lemma}  \newcounter{theorem}
\newcounter{proposition} \newcounter{corol}  \newcounter{remark} \newcounter{example}

\newenvironment{lemma}{\par\refstepcounter{lemma}
     \textbf{Lemma \thelemma.} }{\rm\par}
\newenvironment{theorem}{\par\refstepcounter{theorem}
     \textbf{Theorem \thetheorem.}\ }{\rm\par}
\newenvironment{proposition}{\par\refstepcounter{proposition}
     \textbf{Proposition \theproposition.}\ }{\rm\par}
\newenvironment{corollary}{\par\refstepcounter{corol}
     \textbf{Corollary \thecorol.} }{\rm\par}

\newenvironment{remark}{\par\refstepcounter{remark}
     \textbf{Remark \theremark.}}{\rm\par}

\title{Adaptation of the Alicki-Fannes-Winter method for the set of states with bounded energy and its use.}
\author{ M.E.~Shirokov\thanks{e-mail:msh@mi.ras.ru} \\ Steklov Mathematical Institute, Moscow, Russia}
\date{}

\begin{document}

\maketitle
\begin{abstract}
We describe a  modification of the Alicki-Fannes-Winter method which allows to prove uniform continuity on the set of quantum states with bounded energy of any locally almost affine  function  having limited growth with increasing energy.

Some applications in quantum information theory are considered. The asymptotic  continuity of the relative entropy of entanglement and of its
regularization under the energy constraint on one subsystem is proved. Channel-independent continuity bounds for the quantum mutual information at the output of a local channel and for the output Holevo quantity under the input energy constraint are obtained.
\end{abstract}


\section{Introduction}

R.Alicki and M.Fannes obtained in \cite{A&F} a continuity bound (estimate for variation) for the quantum conditional entropy by using the elegant geometric method based on finding for given quantum states $\rho$ and $\sigma$ such states $\tau_{\rho}$ and $\tau_{\sigma}$ that
\begin{equation}\label{eq-1}
(1-q)\rho+q\tau_{\rho}=(1-q)\sigma+q\tau_{\sigma},
\end{equation}
where $q=q(\|\rho-\sigma\|_1)$ is a number in $[0,1]$ as small as possible vanishing as $\|\rho-\sigma\|_1\rightarrow0$.

B.Synak-Radtke and  M.Horodecki  mentioned in \cite{S&H} that this method can be applied to the class of bounded functions "robust under admixtures".
Then, by using general facts from the state discrimination theory,  M.Mosonyi and F.Hiai  noted in \cite{M&H} that  the states $\tau_{\rho}$ and $\tau_{\sigma}$ chosen  in \cite{A&F} are not optimal and pointed that the optimal states $\tau_{\rho}$ and $\tau_{\sigma}$ in (\ref{eq-1}) are  proportional to the negative part $[\rho-\sigma]_-$ and the positive part $[\rho-\sigma]_+$ of the operator $\rho-\sigma$ correspondingly. They also showed (in the proof of Proposition VI.1 in \cite{M&H}) that the Alicki-Fannes method can be applied to any bounded function $f$ on the set $\S(\H)$ of quantum states which is not "too convex and too concave" in the following sense
\begin{equation}\label{LAA-d}
|f(p\rho+(1-p)\sigma)-p f(\rho)-(1-p)f(\sigma)|\leq r(p)\quad \textup{for all}\;\rho,\sigma\in\S(\H)
\end{equation}
and any $p\in(0,1)$, where $r(p)$ is a vanishing function as $p\rightarrow+0$.\footnote{It means that $\,|f(p\rho+(1-p)\sigma)-p f(\rho)-(1-p)f(\sigma)|\,$ tends to zero as $\,p\rightarrow+0$ uniformly on $\,\S(\H)\times\S(\H)$.} Recently A.Winter proposed additional optimization of the arguments from \cite{M&H}, which makes it possible to obtain \emph{tight} (sharp, in a sense) continuity bounds for the conditional entropy and for the relative entropy of entanglement \cite{W-CB}. In fact, the obtained technique (in what follows we will call it the Alicki-Fannes-Winter method, briefly, the AFW-method) is quite universal, it allows to derive tight  continuity bounds for many entropic and information charateristics of quantum systems and channels \cite{CHI}.

We will call functions $f$ satisfying (\ref{LAA-d}) \emph{locally almost affine}, briefly, LAA\nobreakdash-\hspace{0pt}functions. In quantum information theory the following two classes of LAA\nobreakdash-\hspace{0pt}functions are widely used:
\begin{itemize}
  \item real linear combinations of marginal entropies of a state of a composite quantum system (see Section 5.1);
  \item relative entropy distances from a state to convex sets of states (see Section 5.2).
\end{itemize}

In particular, the AFW-method shows that any locally almost affine bounded function on $\S(\H)$ is uniformly continuous on $\S(\H)$.

The AFW-method can be used regardless of the dimension of the underlying Hilbert space $\H$ under the condition that $f$ is a bounded function on
the whole set of states. But in  analysis of infinite-dimensional quantum systems we often deal with functions which are well defined and bounded only on the sets of states with bounded energy, i.e. states $\rho$ satisfying the inequality
\begin{equation}\label{b-e-ineq}
\Tr H\rho\leq E,
\end{equation}
where $H$ is a positive operator -- the Hamiltonian of a quantum system associated with the space $\H$ \cite{W-CB,H-SCI,H-c-w-c, Wilde+}.

The main obstacle for direct application of the AFW-method to functions on the set of states with bounded energy consists in the difficulty to estimate the energy of the states proportional to the operators $\,[\rho-\sigma]_\pm$  for any states $\rho$ and $\sigma$ satisfying (\ref{b-e-ineq}).
To avoid this problem A.Winter recently proposed the two-step technique based on combination of the AFW-method with the special finite-dimensional approximation of arbitrary states with bounded energy \cite{W-CB}. This technique allows to obtain asymptotically tight continuity bounds for the von Neumann entropy, for the conditional entropy and for the conditional mutual information  under the constraint (\ref{b-e-ineq}) on one subsystem \cite{W-CB,CHI}. But application of Winter's technique to any LAA-function $f$  is limited by the approximation step, since it requires special estimates depending on $f$.

In the first part of the paper we propose a modification of the AFW-method which can be applied \emph{directly} (without approximation) to any LAA-function $f$ such that
\begin{equation}\label{g-cond}
  \sup_{\Tr H\rho\leq E}|f(\rho)|=o\shs(\sqrt{E})\quad\textup{as}\quad E\rightarrow+\infty.
\end{equation}
The main idea of this modification is using the operators $\,\Tr_R\,[\hat{\rho}-\hat{\sigma}]_\pm$, where $\hat{\rho}$ and $\hat{\sigma}$ are appropriate purifications of given states $\rho$ and $\sigma$ satisfying (\ref{b-e-ineq}).

Condition (\ref{g-cond}) is essential (note that the affine function $\rho\mapsto\Tr H\rho$ may be discontinuous on the set of states satisfying (\ref{b-e-ineq})). Fortunately, this condition is valid for many entropic characteristics of states of a quantum system provided the Hamiltonian $H$ satisfies the condition
\begin{equation}\label{H-cond+p}
  \lim_{\lambda\rightarrow+0}\left[\Tr e^{-\lambda H}\right]^{\lambda}=1,
\end{equation}
which holds, in particular, for the system of quantum oscillators playing a central role in continuous variable quantum information theory \cite{H-SCI,W&Co}.

In the second  part of the paper we consider several applications of the proposed modification of the AFW-method. In particular, we
obtain  continuity bounds for the relative entropy of entanglement and for its
regularization under the energy constraint on one subsystem which imply the asymptotic continuity of these characteristics (if the Hamiltonian of the subsystem satisfies condition (\ref{H-cond+p})). We also obtain channel-independent continuity bounds for the quantum mutual information at the output of a local channel and for the output Holevo quantity under the input energy constraint.

\section{Preliminaries}

Let $\mathcal{H}$ be a separable infinite-dimensional Hilbert space,
$\mathfrak{B}(\mathcal{H})$ the algebra of all bounded operators on $\mathcal{H}$ with the operator norm $\|\cdot\|$ and $\mathfrak{T}( \mathcal{H})$ the
Banach space of all trace-class
operators   with the trace norm $\|\!\cdot\!\|_1$. Let
$\mathfrak{S}(\mathcal{H})$ be  the set of quantum states (positive operators
in $\mathfrak{T}(\mathcal{H})$ with unit trace) \cite{H-SCI,N&Ch,Wilde}.

Denote by $I_{\mathcal{H}}$ the identity operator in a Hilbert space
$\mathcal{H}$ and by $\id_{\mathcal{\H}}$ the identity
transformation of the Banach space $\mathfrak{T}(\mathcal{H})$.

If quantum systems $A$ and $B$ are described by Hilbert spaces  $\H_A$ and $\H_B$ then the bipartite system $AB$ is described by the tensor product of these spaces, i.e. $\H_{AB}\doteq\H_A\otimes\H_B$. A state in $\S(\H_{AB})$ is denoted $\rho_{AB}$, its marginal states $\Tr_{B}\rho_{AB}\doteq\Tr_{\H_B}\rho_{AB}$ and $\Tr_{A}\rho_{AB}\doteq\Tr_{\H_A}\rho_{AB}$ are denoted respectively $\rho_{A}$ and $\rho_{B}$.

The \emph{von Neumann entropy} $H(\rho)=\mathrm{Tr}\eta(\rho)$ of a
state $\rho\in\mathfrak{S}(\mathcal{H})$, where $\eta(x)=-x\log x$,
is a concave nonnegative lower semicontinuous function on the set $\mathfrak{S}(\mathcal{H})$ \cite{H-SCI,L-2}.
The concavity of the von Neumann entropy is supplemented by the
inequality
\begin{equation}\label{w-k-ineq}
H(p\rho+(1-p)\sigma)\leq pH(\rho)+(1-p)H(\sigma)+h_2(p),
\end{equation}
where $h_2(p)=\eta(p)+\eta(1-p)$, valid for any
states  $\rho,\sigma\in\S(\H)$ and $p\in(0,1)$ \cite{H-SCI,N&Ch}.

The \emph{quantum conditional entropy}
\begin{equation*}
H(A|B)_{\rho}=H(\rho_{AB})-H(\rho_B)
\end{equation*}
of a bipartite state $\rho_{AB}$ with finite marginal entropies is essentially used in analysis of quantum systems \cite{H-SCI,Wilde}. The conditional entropy is concave and satisfies the following inequality
\begin{equation}\label{ce-ac}
H(A|B)_{p\rho+(1-p)\sigma}\leq p H(A|B)_{\rho}+(1-p)H(A|B)_{\sigma}+h_2(p)
\end{equation}
for any  $p\in(0,1)$ and any states $\rho_{AB}$ and $\sigma_{AB}$. Inequality (\ref{ce-ac}) follows from concavity of the entropy and  inequality (\ref{w-k-ineq}).

The \emph{quantum relative entropy} for two states $\rho$ and
$\sigma$ in $\mathfrak{S}(\mathcal{H})$ is defined by the formula
$$
H(\rho\shs\|\shs\sigma)=\sum_i\langle
i|\,\rho\log\rho-\rho\log\sigma\,|i\rangle,
$$
where $\{|i\rangle\}$ is the orthonormal basis of
eigenvectors of the state $\rho$ and it is assumed that
$H(\rho\shs\|\shs\sigma)=+\infty$ if $\,\mathrm{supp}\rho\shs$ is not
contained in $\shs\mathrm{supp}\shs\sigma$ \cite{H-SCI,L-2}.\footnote{The support $\,\mathrm{supp}\rho\,$ of a positive operator $\rho$ is the orthogonal complement to its kernel.}

The \emph{quantum mutual information} of a state $\,\rho_{AB}\,$ of a
bipartite quantum system  is defined as
\begin{equation*}
I(A\!:\!B)_{\rho}=H(\rho_{AB}\shs\Vert\shs\rho_{A}\otimes
\rho_{B})=H(\rho_{A})+H(\rho_{B})-H(\rho_{AB}),
\end{equation*}
where the second expression  is valid if $\,H(\rho_{AB})\,$ is finite \cite{L-mi}.

Basic properties of the relative entropy show that $\,\rho\mapsto
I(A\!:\!B)_{\rho}\,$ is a lower semicontinuous function on the set
$\S(\H_{AB})$ taking values in $[0,+\infty]$. It is well known that
\begin{equation}\label{MI-UB}
I(A\!:\!B)_{\rho}\leq 2\min\left\{H(\rho_A),H(\rho_B)\right\}
\end{equation}
for any state $\rho_{AB}$ \cite{L-mi,MI-B}.

The quantum mutual information is not concave or convex but the inequality
\begin{equation}\label{F-c-b}
\begin{array}{cc}
\left|p
I(A\!:\!B)_{\rho}+(1-p)I(A\!:\!B)_{\sigma}-I(A\!:\!B)_{p\rho+(1-p)\sigma}\right|\leq h_2(p)
\end{array}
\end{equation}
holds for $p\in(0,1)$ and any states $\rho_{AB}$, $\sigma_{AB}$ with finite $I(A\!:\!B)_{\rho}$, $I(A\!:\!B)_{\sigma}$.
If $\rho_{AB}$, $\sigma_{AB}$ are states with finite marginal entropies then (\ref{F-c-b}) can be easily proved by noting that
\begin{equation}\label{I-rep}
I(A\!:\!B)_{\rho}=H(\rho_A)-H(A|B)_{\rho},
\end{equation}
and by using the concavity of the entropy and of the conditional entropy along with the inequalities
(\ref{w-k-ineq}) and (\ref{ce-ac}). The validity of inequality (\ref{F-c-b}) for any states $\rho_{AB}$, $\sigma_{AB}$  with finite mutual information can be proved by approximation (using Theorem 1 in \cite{CMI}).

\section{Basic results}

Let $H$ be a positive operator on a Hilbert space $\H$ and $E\geq E_0\doteq\inf\limits_{\|\varphi\|=1}\langle\varphi|H|\varphi\rangle$. Then
$$
\E_{H,E}=\{\rho\in\S(\H)\,|\,\Tr H\rho\leq E\shs\}
$$
is a closed convex subset of $\S(\H)$.\footnote{The value of $\,\Tr H\rho$ (finite or infinite) is defined as $\,\sup_n \Tr P_n H\rho$, where $P_n$ is the spectral projector of $H$ corresponding to the interval $[0,n]$.} If $H$ is the Hamiltonian of the quantum system associated with the space $\H$ then
$\E_{H,E}$ is the set of states with mean energy not exceeding $E$.

Let $f$ be a function defined on the set $\E_{H,\infty}\doteq\bigcup_{E\geq E_0}\E_{H,E}$. We will say that $f$ is  \emph{locally almost affine} function, briefly, LAA-function if
\begin{equation}\label{a-a-f}
-a(p)\leq f(p\rho+(1-p)\sigma)-p f(\rho)-(1-p)f(\sigma)\leq b(p)
\end{equation}
for any $p\in(0,1)$ and all $\rho,\sigma\in\E_{H,\infty}$, where $a(p)$ and $b(p)$ are nonnegative functions on $(0,1)$ vanishing as $p\rightarrow+0$.\smallskip

\begin{theorem}\label{AFM} \emph{If $\,f$ is a function on the set $\,\E_{H,\infty}$ possessing property (\ref{a-a-f}) such that $\,B_f(E)\doteq\sup\limits_{\rho\in\E_{H,E}}|f(\rho)|<+\infty\,$  for all finite $\,E\geq E_0\,$ then
\begin{equation}\label{f-CB}
|f(\rho)-f(\sigma)|\leq 2\sqrt{2\varepsilon}B_f\!\left(E/\varepsilon\right)+(1+\sqrt{2\varepsilon})(a(\epsilon)+b(\epsilon)),
\end{equation}
where  $\,\epsilon=\sqrt{2\varepsilon}/(1+\sqrt{2\varepsilon})$, for any states $\rho$ and $\sigma$ in $\,\E_{H,E}$ such that $\,\frac{1}{2}\|\rho-\sigma\|_1\leq\varepsilon\leq \frac{1}{2}$. The term $\,2B_f\!\left(E/\varepsilon\right)$ in the right hand side of (\ref{f-CB}) can be replaced by $\,B^+_f\!\left(E/\varepsilon\right)+B^-_f\!\left(E/\varepsilon\right)$, where $B^\pm_f(E)\doteq\sup\limits_{\rho\in\E_{H,E}} \max\{\pm f(\rho),0\}$.}

\emph{For pure states $\rho$ and $\sigma$ inequality (\ref{f-CB}) holds with $\shs\varepsilon$ replaced by $\,\varepsilon^2/2$.}
\end{theorem}\smallskip

\begin{remark}\label{AFM-r} We assume that  $\,\frac{1}{2}\|\rho-\sigma\|_1\leq\varepsilon$  (instead of $\,\frac{1}{2}\|\rho-\sigma\|_1=\varepsilon$), since we can not guarantee, in general, that the right hand side of (\ref{f-CB}) is a nondecreasing function of $\varepsilon$ even in the case when it tends to zero as $\varepsilon\rightarrow0$.
\end{remark}\smallskip

\begin{corollary}\label{AFM-c} \emph{If $\,f$ is a LAA-function on $\,\C_{H,\infty}$ such that $B_f(E)=o(\sqrt{E})$ as $E\rightarrow+\infty$ then $f$ is uniformly continuous on the set
$\,\C_{H,E}$ for any finite $E\geq E_0$.}
\end{corollary}\smallskip

\emph{Proof of Theorem 1.} Let $\H_R\cong\H$. Since $\,\frac{1}{2}\|\rho-\sigma\|_1\leq\varepsilon\leq\frac{1}{2}$, in $\S(\H\otimes\H_R)$ there exist purifications  $\hat{\rho}=|\varphi\rangle\langle\varphi|$ and  $\hat{\sigma}=|\psi\rangle\langle\psi|$ of the states $\rho$ and $\sigma$  such that $\delta\doteq\frac{1}{2}\|\hat{\rho}-\hat{\sigma}\|_1=\sqrt{2\varepsilon}$ \cite{H-SCI,Wilde}. Note that $\delta=\sqrt{1-|\langle\varphi|\psi\rangle|^2}$.

Following \cite{M&H,W-CB} introduce the quantum states
$\,\hat{\tau}_+=\delta^{-1}[\shs\hat{\rho}-\hat{\sigma}\shs]_+\,$
and
$\,\hat{\tau}_-=\delta^{-1}[\shs\hat{\rho}-\hat{\sigma}\shs]_-\,$ such that
\begin{equation*}
\frac{1}{1+\delta}\,\hat{\rho}+\frac{\delta}{1+\delta}\,\hat{\tau}_-=\omega_*=
\frac{1}{1+\delta}\,\hat{\sigma}+\frac{\delta}{1+\delta}\,\hat{\tau}_+.
\end{equation*}
By taking partial trace we obtain
\begin{equation}\label{omega-star}
\frac{1}{1+\delta}\,\rho+\frac{\delta}{1+\delta}\,\tau_-=\Tr_{\!R}\shs\omega_*=
\frac{1}{1+\delta}\,\sigma+\frac{\delta}{1+\delta}\,\tau_+,
\end{equation}
where $\tau_{\pm}=\Tr_R\shs\hat{\tau}_{\pm}$.

By using spectral decomposition of the operator $\;\hat{\rho}-\hat{\sigma}=|\varphi\rangle\langle\varphi|-|\psi\rangle\langle\psi|$ one can show that
$\hat{\tau}_{\pm}$ are pure states corresponding to the unit vectors
$$
|\gamma_{\pm}\rangle=p_{\pm}|\varphi\rangle+q_{\pm}|\psi\rangle,\;\,\textrm{where}\;\;p_{\pm}=\frac{\langle\varphi|\psi\rangle}{\delta\sqrt{2(1\mp\delta)}},\;\;
q_{\pm}=-\frac{(1\mp\delta)}{\delta\sqrt{2(1\mp\delta)}}.
$$
So, we have
$$
\begin{array}{c}
\Tr H\tau_{\pm}=\langle\gamma_{\pm}|H\otimes I_R|\gamma_{\pm}\rangle=|p_{\pm}|^2\langle\varphi|H\otimes I_R|\varphi\rangle+|q_{\pm}|^2\langle\psi|H\otimes I_R|\psi\rangle
\\\\+2\Re\shs \bar{p}_{\pm}q_{\pm}\langle\varphi|H\otimes I_R|\psi\rangle\leq |p_{\pm}|^2\Tr H\rho+|q_{\pm}|^2\Tr H\sigma+2|p_{\pm}q_{\pm}|\sqrt{\Tr H\rho}\sqrt{\Tr H\sigma}\\\\\leq E(|p_{\pm}|+|q_{\pm}|)^2=(1+|\langle\varphi|\psi\rangle|)E/\delta^2\leq 2E/\delta^2=E/\varepsilon,
\end{array}
$$
where the Schwarz inequality was used.

It follows that the states $\tau_{\pm}$ belong to the set $\C_{H,E/\varepsilon}$  and hence
\begin{equation}\label{f-UB}
|f(\tau_{\pm})|\leq B_f(E/\varepsilon).
\end{equation}
By applying (\ref{a-a-f}) to the convex decompositions of the state $\Tr_{\!R}\shs\omega_*$ in (\ref{omega-star}) we obtain
$$
(1-p)[f(\rho)-f(\sigma)]\leq p
[f(\tau_+)-f(\tau_-)]+a(p)+b(p)
$$
and
$$
(1-p)[f(\sigma)-f(\rho)]\leq p
[f(\tau_-)-f(\tau_+)]+a(p)+b(p)
$$
where $p=\frac{\delta}{1+\delta}$. These inequalities and  upper bound (\ref{f-UB}) imply inequality (\ref{f-CB}).
Since $\,|f(\tau_+)-f(\tau_-)|\leq B^+_f(E/\varepsilon)+B^-_f(E/\varepsilon)$, the term $\,2B_f\left(E/\varepsilon\right)$ in (\ref{f-CB}) can be replaced by $\,B^+_f\left(E/\varepsilon\right)+B^-_f\left(E/\varepsilon\right)$.

If $\rho$ and $\sigma$ are pure states then we can take pure states $\hat{\rho}=\rho\otimes\varrho$ and  $\hat{\sigma}=\sigma\otimes\varsigma$ such that $\frac{1}{2}\|\hat{\rho}-\hat{\sigma}\|_1=\varepsilon$ and repeat the above arguments. \rule{5pt}{5pt}\smallskip

\begin{remark}\label{AFM-r} In applications we often deal with a function $f$ which is defined and locally almost affine on the set $\C^0_{H,\infty}\doteq\bigcup_E\C^0_{H,E}$, where $\C^0_{H,E}$ is a convex subset of $\C_{H,E}$ for each $E$ (for example, $\C^0_{H,E}$ is the subset of $\C_{H,E}$ consisting of finite rank states, etc.). The proof of Theorem \ref{AFM} shows that its assertion is valid for $\C^0_{H,E}$ instead of $\C_{H,E}$ if the following condition holds:
\begin{equation}\label{sp-cond}
  \textrm{the states }\,c^{-1}_{\pm}\Tr_{R}\,[\hat{\rho}-\hat{\sigma}]_{\pm}\,\textrm{ belong to the set }\C^0_{H,\infty},
\end{equation}
where $c_{\pm}=\Tr[\hat{\rho}-\hat{\sigma}]_{\pm}$, for any purifications $\hat{\rho}$ and $\hat{\sigma}$ in $\S(\H\otimes\H_R)$ of arbitrary states $\rho$ and $\sigma$ in $\C^0_{H,E}$.
\end{remark}

\begin{corollary}\label{AFM-c+} \emph{Let $\,\C^0_{H,E}$ be a dense subset of $\,\C_{H,E}$ for each $E\geq E_0$ such that condition (\ref{sp-cond}) holds. If $\,f$ is a LAA-function on $\,\C^0_{H,\infty}$ such that
$$
B_f(E)\doteq\sup_{\rho\in\C^{0}_{H,E}}|f(\rho)|=o\shs(\sqrt{E})\;\textit{ as }\;E\rightarrow+\infty
$$ then $f$ has a uniformly continuous extension  to the set
$\,\C_{H,E}$ for any finite $E\geq E_0$ satisfying (\ref{f-CB}).}
\end{corollary}

\section{Functions majorized by a marginal entropy}

In this section we specify the universal results of Section 3 for special class of functions used in quantum information theory.

\subsection{General case}

Many important characteristics of states of a finite-dimensional $n$-partite
system $A_{1}...A_{n}$ have a form of a function $f$ on the set $\S(\H_{A_{1}...A_{n}})$
satisfying the inequalities
\begin{equation}\label{F-p-1}
-a_f h_2(p)\leq f(p\rho+(1-p)\sigma)-p f(\rho)-(1-p)f(\sigma)\leq b_fh_2(p),
\end{equation}
where $p\in(0,1)$, $h_2$ is the binary entropy (defined after (\ref{w-k-ineq})), $\,a_f\,b_f\in \mathbb{R}_+$,
and
\begin{equation}\label{F-p-2}
-c^-_f H(\rho_{B})\leq f(\rho)\leq c^+_fH(\rho_{B}),
\end{equation}
where $B$ is a particular subsystem of $A_{1}...A_{n}$ and $c^-_f,c^+_f\in \mathbb{R}_+$.
Examples of  characteristics satisfying (\ref{F-p-1}) and  (\ref{F-p-2}) are considered in Sections 5.1 and 5.2.2.

To formulate the main result of this section consider the function
\begin{equation}\label{F-def}
F_{H_B}(E)\doteq\sup_{\Tr H_B\rho\leq E}H(\rho),\quad E\geq E_0\doteq\inf\limits_{\|\varphi\|=1}\langle\varphi|H_B|\varphi\rangle,
\end{equation}
where $H_B$ is the Hamiltonian of the system $B$ (involved in (\ref{F-p-2})).
Properties of this function are described in Proposition 1 in \cite{EC}. It shows, in particular, that
\begin{equation}\label{F-exp}
F_{H_B}(E)=\lambda(E)E+\log\Tr e^{-\lambda(E) H_{\!B}}=o(E)\quad\textup{as}\quad E\rightarrow+\infty,
\end{equation}
where $\lambda(E)$ is determined by the equality $\Tr H_B e^{-\lambda(E) H_B}=E\Tr e^{-\lambda(E) H_B}$, provided that
\begin{equation}\label{H-cond}
  \Tr e^{-\lambda H_{B}}<+\infty\quad\textrm{for all}\;\lambda>0.
\end{equation}
It is well known that condition (\ref{H-cond}) implies  continuity of the von Neumann entropy on the set $\C_{H_B\!,E}$ for any $E\geq E_0$ and  attainability of the supremum in (\ref{F-def})  at the \emph{Gibbs state} $\gamma_B(E)\doteq e^{-\lambda(E) H_B}/\Tr e^{-\lambda(E) H_B}$ \cite{W}. So, we have $F_{H_B}(E)=H(\gamma_B(E))$ for any $E\geq E_0$.\smallskip

Note also that condition (\ref{H-cond}) implies that the operator $H_B$ has a discrete spectrum of finite multiplicity, i.e. it can be represented as $$
H_B=\sum_{k=0}^{+\infty}E_k|\tau_k\rangle\langle \tau_k|,
$$
where $\{E_k\}$ is the nondecreasing sequence of eigenvalues of $H_B$ tending to $+\infty$ and $\{|\tau_k\rangle\}$ -- the corresponding basis of eigenvectors.

To apply  the modified AFW-method to functions satisfying (\ref{F-p-1}) and (\ref{F-p-2}) one has to slightly strengthen condition (\ref{H-cond}).\pagebreak

\begin{proposition}\label{SCB-1} \emph{Let $f$ be a function on the set $\,\{\shs\rho_{A_{1}...A_{n}}\shs|\,\Tr H_B\rho_{B}<+\infty\shs\}$  satisfying (\ref{F-p-1}) and (\ref{F-p-2}). Then
\begin{equation}\label{SBC-ineq}
    |f(\rho)-f(\sigma)|\leq (c^-_f+c^+_f)\sqrt{2\varepsilon} F_{H_B}\!\left(E/\varepsilon\right)+(a_f+b_f)g(\sqrt{2\varepsilon})
\end{equation}
for any states $\shs\rho\shs$ and $\shs\sigma\shs$  such that $\;\Tr H_B\rho_{B},\Tr H_B\sigma_{B}\leq E\,$ and $\;\,\frac{1}{2}\|\shs\rho-\sigma\|_1\leq\varepsilon\leq\frac{1}{2}$, where $F_{H_B}$ is the function defined in (\ref{F-def}) and  $\,g(x)\!\doteq\!(1+x)h_2\!\left(\frac{x}{1+x}\right)$.}

\emph{For pure states $\rho$ and $\sigma$ inequality (\ref{SBC-ineq}) holds with $\,\varepsilon$ replaced by $\,\varepsilon^2/2$.}\smallskip

\emph{The right hand side of  (\ref{SBC-ineq}) tends to zero as $\,\varepsilon\rightarrow0\,$ if and only if the Hamiltonian  $H_B$  satisfies the condition
\begin{equation}\label{H-cond+}
  \lim_{\lambda\rightarrow+0}\left[\Tr e^{-\lambda H_B}\right]^{\lambda}=1.
\end{equation}
If this condition holds then the function $f$ is uniformly continuous on the set
$\,\{\shs\rho_{A_{1}...A_{n}}\shs|\,\Tr H_B\rho_{B}\leq E\shs\}$ for any  $E\geq E_0$.}\smallskip

\emph{Condition (\ref{H-cond+}) holds if  the Hamiltonian $H_B$ has the discrete spectrum $\{E_k\}_{k\geq0}$ such that $\;\liminf\limits_{k\rightarrow\infty} E_k/\log^q k>0\,$ for some $\,q>2$.}\footnote{By Lemma \ref{SCB-l-2} below condition (\ref{H-cond+}) is not valid if $\;\limsup\limits_{k\rightarrow\infty} E_k/\log^2 k<+\infty$.}
\end{proposition}

\begin{remark}\label{SCB-r} Condition (\ref{H-cond+}) is stronger than condition (\ref{H-cond}). By Proposition 1 in \cite{EC} and Lemma \ref{SCB-l-1} below these conditions can be written in terms of the function $F_{H_B}$ as $F_{H_B}(E)=o\shs(\!\sqrt{E})$ and $F_{H_B}(E)=o(E)$ for large $E$ correspondingly. In terms of the sequence $\{E_k\}$ of eigenvalues of $H_B$
condition (\ref{H-cond}) means that $\lim_kE_k/\log k=+\infty$. Hence,  the last assertion of Proposition \ref{SCB-1} shows that the difference between conditions (\ref{H-cond}) and (\ref{H-cond+}) is not too large. It is essential that condition
(\ref{H-cond+}) holds for the Hamiltonian of the  system of quantum oscillators (see the next subsection).
\end{remark}

We will use the following two lemmas proved in the Appendix.\smallskip

\begin{lemma}\label{SCB-l-1} \emph{Condition (\ref{H-cond+}) is equivalent to the following one}
$$
F_{H_B}(E)=o\shs(\!\sqrt{E})\quad \textit{as} \quad E\rightarrow\!+\infty.
$$
\end{lemma}

\begin{lemma}\label{SCB-l-2} \emph{Let $\,E_k=\log^q k$, $k=1,2,...$, then
$\lim\limits_{\lambda\rightarrow+0}\left[\sum_{k\geq 1} e^{-\lambda E_k}\right]^{\lambda}=1$ if and only if $\,q>2$.}
\end{lemma}\smallskip

\emph{Proof of Proposition \ref{SCB-1}.} Let $\bar{B}=A_{1}...A_{n}\setminus B$ and $\hat{H}=H_B\otimes I_{\bar{B}}$ be a positive operator in $\H_{A_{1}...A_{n}}$. Then $\,\{\shs\rho_{A_{1}...A_{n}}\shs|\,\Tr H_B\rho_{B}\leq E\shs\}=\C_{\hat{H},E}$ in terms of Section 3.  So, the main assertions of the proposition follow from Theorem \ref{AFM} and Lemma \ref{SCB-l-1}.

The last  assertion  follows from Lemma \ref{SCB-l-2}, since it is easy to see that
$$
\lim_{\lambda\rightarrow+0}\left[\sum_{k=0}^{+\infty} e^{-\lambda E_k}\right]^{\lambda}=1\quad \Leftrightarrow \quad\lim_{\lambda\rightarrow+0}\left[\sum_{k=n}^{+\infty} e^{-\lambda E_k}\right]^{\lambda}=1
$$
for any sequence $\{E_k\}$ of positive numbers and any given $n$. \rule{5pt}{5pt}

\subsection{The $\ell$-mode quantum oscillator}

Consider now the case when the system $B$ in (\ref{F-p-2}) is the $\ell$-mode quantum oscillator. In this case
$$
H_B=\sum_{i=1}^{\ell}\hbar\shs\omega_i\!\left(a^{\dagger}_ia_i+\textstyle\frac{1}{2}I_B\right),
$$
where $\,a_i\,$ and $\,a^{\dagger}_i\,$ are the annihilation and creation operators and $\,\omega_i\,$ is the frequency of the $i$-th oscillator \cite[Ch.12]{H-SCI}. It follows that\vspace{-5pt}
$$
F_{H_B}(E)=\max_{\{E_i\}}\sum_{i=1}^{\ell}g\!\left(E_i/\hbar\omega_i-1/2\right),\quad E\geq E_0\doteq\frac{1}{2}\sum_{i=1}^{\ell}\hbar\omega_i,\vspace{-5pt}
$$
where $\,g(x)=(x+1)\log(x+1)-x\log x\,$ and the maximum is over all $\ell\textup{-}$tuples $E_1$,...,$E_{\ell}$  such that  $\sum_{i=1}^{\ell}E_i=E$ and $E_i\geq\frac{1}{2}\hbar\omega_i$. The exact value of
$F_{H_B}(E)$ can be calculated by applying the Lagrange multiplier method which leads to a transcendental equation. But following \cite{W-CB} one can obtain
upper bound for $F_{H_B}(E)$ by using the inequality $\,g(x)\leq\log(x+1)+1\,$  valid for all $\,x>0\,$. It implies
$$
F_{H_B}(E)\leq \max_{\sum_{i=1}^{\ell}E_i=E}\sum_{i=1}^{\ell}\log\left(E_i/\hbar\omega_i+1/2\right)+\ell.
$$
By calculating this maximum via the Lagrange multiplier method  we obtain
\begin{equation}\label{bF-ub}
F_{H_B}(E)\leq \widehat{F}_{\ell,\omega}(E)\doteq\ell\log \frac{E+E_0}{\ell E_*}+\ell,\quad E_*=\left[\prod_{i=1}^{\ell}\hbar\omega_i\right]^{1/\ell}.\vspace{-5pt}
\end{equation}
It is easy to see that upper bound (\ref{bF-ub}) is $\varepsilon$-sharp for large $E$. By using this upper bound one can derive from Proposition \ref{SCB-1} the following\smallskip

\begin{corollary}\label{SCB-G-1}
\emph{Let $f$ be a function on the set $\,\{\shs\rho_{A_{1}...A_{n}}\shs|\,\Tr H_B\rho_{B}<+\infty\shs\}$  satisfying (\ref{F-p-1}) and (\ref{F-p-2})
in which $B$ is the $\ell$-mode quantum oscillator with the frequencies $\,\omega_1,...,\omega_{\ell}$,
$\,E>E_0\doteq\frac{1}{2}\sum_{i=1}^{\ell}\hbar\omega_i\,$ and $\,E_*=[\prod_{i=1}^{\ell}\hbar\omega_i]^{1/\ell}$. Then
\begin{equation}\label{SBC-ineq-G}
   \! |f(\rho)-f(\sigma)|\leq (c^-_f+c^+_f)\sqrt{2\varepsilon}\ell\!\left(\log \frac{E/\varepsilon+E_0}{\ell E_*}+1\!\right)+(a_f+b_f)g(\sqrt{2\varepsilon})
\end{equation}
for any states $\rho$ and $\sigma$ such that $\,\Tr H_B\rho_{B},\Tr H_B\sigma_{B}\leq E$ and $\;\frac{1}{2}\|\shs\rho-\sigma\|_1\leq\varepsilon\leq\frac{1}{2}$.}\smallskip

\emph{For pure states $\rho$ and $\sigma$ inequality (\ref{SBC-ineq-G}) holds with $\,\varepsilon$ replaced by $\,\varepsilon^2/2$.}
\end{corollary}

\section{Applications}

\subsection{Linear combinations of marginal entropies}

Several important entropic characteristics of a state  of a finite-dimensional $n$-partite
system $A_{1}...A_{n}$  are defined as a real linear combination of marginal
entropies, i.e. as the function
\begin{equation}\label{F-form}
  f(\rho_{A_{1}...A_{n}})=\sum_k c_k H(\rho_{X_k})
\end{equation}
on the set of all states of the system, where $\rho_{X_k}$ is the
partial state of $\rho_{A_{1}...A_{n}}$ corresponding to the
subsystem $X_k$ of $A_{1}...A_{n}$ and $c_k\in\mathbb{R}$.

By using concavity of the von Neumann entropy and inequality (\ref{w-k-ineq}) it is easy to show that the function $f$  in (\ref{F-form}) satisfies the LAA-property (\ref{F-p-1}) with $\,a_f\leq\sum_{k:c_k<0}|c_k|$  and $\,b_f\leq\sum_{k:c_k>0}c_k$.\footnote{Inequality (\ref{F-c-b}) shows that the coefficients $a_f$ and $b_f$ may be  less than $\sum_{k:c_k<0}|c_k|$ and $\sum_{k:c_k>0}c_k$.}

It is also essential that many important characteristics  having form (\ref{F-form}) possess lower and upper bounds proportional to one of the marginal entropies, i.e. they satisfy the inequality (\ref{F-p-2}) for a particular subsystem $B$ of $A_{1}...A_{n}$ and some nonnegative numbers $c^-_f,c^+_f$. For example, the quantum mutual information
$I(A_1\!:\!A_2)_{\rho}$ considered as a function  of a state $\rho_{A_1A_2A_3}$ is nonnegative and upper bounded by one of the quantities:
$$
2H(\rho_{A_1}),\;2H(\rho_{A_2}),\;2H(\rho_{A_1A_3}),\;2H(\rho_{A_2A_3}).
$$
This follows from the inequality
$I(A\!:\!B)\leq I(A\!:\!BC)$ and  upper bound (\ref{MI-UB}).

In finite dimensions the properties (\ref{F-p-1}) and (\ref{F-p-2}) make it possible to directly apply the AFW-method to the function $f$ and obtain the continuity bound
\begin{equation}\label{FA-ineq}
    |f(\rho)-f(\sigma)|\leq (c^-_f+c^+_f)\shs\varepsilon\log\dim\H_B+(a_f+b_f)g(\varepsilon),
\end{equation}
where $\;\varepsilon=\frac{1}{2}\|\shs\rho-\sigma\|_1\,$ and $\,g(\varepsilon )\!\doteq\!(1+\varepsilon)h_2\!\left(\frac{\varepsilon}{1+\varepsilon}\right)$ \cite[Proposition 1]{CMI}.

By using (\ref{FA-ineq}) and Winter's technique from \cite{W-CB}  based on a finite-dimensional approximation one can obtain continuity bounds for several characteristics having form (\ref{F-form}), in particular, for the von Neumann entropy, the conditional entropy and the conditional mutual information  under the energy constraint on one subsystem \cite{W-CB,CHI}. But application of this technique to arbitrary function (\ref{F-form}) with properties (\ref{F-p-1}) and (\ref{F-p-2})  is limited by the approximation step. The modified AFW-method considered in Sections 3,4 makes it possible to obtain universal continuity bounds for such functions.

In infinite dimensions the right hand side of (\ref{F-form}) is correctly defined
if all the marginal entropies $\,H(\rho_{X_k})\,$ are finite (or at
least the linear combination in (\ref{F-form}) does not contain the uncertainty
$"\infty-\infty"$). So, the function $f$ in (\ref{F-form}) is well defined on the
 dense convex subset
\begin{equation}\label{f-r-s}
  \left\{\shs\rho_{A_{1}...A_{n}}\shs|\,\rank\rho_{A_k}<+\infty,\; k=\overline{1,n}\shs\right\}
\end{equation}
of $\S(\H_{A_{1}...A_{n}})$. Following \cite{CMI} we will say that $f_\mathrm{e}$ is a \emph{faithful} extension of a function $f$ defined on set (\ref{f-r-s}) to the set
$\,\mathcal{B}=\left\{\shs\rho_{A_{1}...A_{n}}\shs|\,H(\rho_{B})<+\infty\shs\right\}\,$ if $f_\mathrm{e}$ coincides with $f$ on set (\ref{f-r-s}) and for arbitrary state $\rho\in\mathcal{B}$ the following property holds:
 \begin{equation*}
\lim_{k\rightarrow\infty}f_\mathrm{e}(\rho^k_{A_{1}...A_{n}})=f_\mathrm{e}(\rho_{A_{1}...A_{n}})\in[-\infty,+\infty]
\end{equation*}
for \emph{any} sequence of "truncated" states
$$
\rho^k_{A_{1}...A_{n}}=\lambda_k^{-1}Q_k\rho_{A_{1}...A_{n}}Q_k,\quad
Q_k= P^k_{A_1}\otimes \ldots\otimes P^k_{A_n},\;\lambda_k=\Tr
Q_k\rho_{A_{1}...A_{n}},
$$
determined by sequences
$\{P^k_{A_{1}}\}_k\subset\B(\H_{A_{1}})$,...,
$\{P^k_{A_{n}}\}_k\subset\B(\H_{A_{n}})$ of  projectors strongly
converging to the unit operators
$I_{A_{1}}$,...,$I_{A_{n}}$.\footnote{Basic properties of the entropy imply that all the states $\rho^k_{A_{1}...A_{n}}$ belong to the set $\mathcal{B}$.}
\smallskip

For example, the conditional entropy $\,H(A_1|A_2)=H(\rho_{A_1A_2})-H(\rho_{A_2})\,$  has the faithful extension
\begin{equation}\label{ce-ext}
H_\mathrm{e}(A_1|A_2)_{\rho}\doteq H(\rho_{A_1})-I(A_1\!:\!A_2)_{\rho}
\end{equation}
to the set
$\,\left\{\shs\rho_{A_{1}A_{2}}\shs|\,H(\rho_{A_1})<+\infty\shs\right\}$ (containing states $\rho_{A_1A_2}$ such that $H(\rho_{A_1A_2})=H(\rho_{A_2})=+\infty$) introduced by Kuznetsova in \cite{Kuz} and studied  in \cite[Section 5]{CMI}.

The expression $I(A_1\!:\!A_2)_{\rho}=H(\rho_{A_1A_2}\shs\Vert\shs\rho_{A_1}\otimes
\rho_{A_2})$ for the quantum mutual information can be considered as a faithful extension  of the linear combination $H(\rho_{A_1})+H(\rho_{A_2})-H(\rho_{A_1A_2})$ to the set $\S(\H_{A_1A_2})$. Faithful extensions of
several other important characteristics having form (\ref{F-form}) and general methods for construction of such extensions can be found in \cite{CMI}.
\smallskip

\begin{proposition}\label{LCME} \emph{Let $f$ be a function having form (\ref{F-form}) such that inequalities  (\ref{F-p-1}) and (\ref{F-p-2}) hold on set (\ref{f-r-s}).  If there is a faithful extension of $f$ to the set $\mathcal{B}$ (of states with finite $H(\rho_B)$) and the Hamiltonian  $H_B$ of the system $B$ in (\ref{F-p-2}) satisfies condition
(\ref{H-cond+}) then  this extension is uniformly continuous on the set
$\C_{H_B\!,E}\doteq\{\shs\rho_{A_{1}...A_{n}}\shs|\,\Tr H_B\rho_{B}\leq E\shs\}$ for any $E>E_0$ and satisfies  continuity bound  (\ref{SBC-ineq}).}

\emph{If $\,B$ is the $\ell$-mode quantum oscillator then the above extension  satisfies continuity bound (\ref{SBC-ineq-G}).}
\end{proposition}\smallskip

\emph{Proof.} Condition (\ref{H-cond+}) implies  that $\,\C_{H_B\!,E}\subset\mathcal{B}\,$ for any $E>E_0$ \cite{EC,W}.

Since inequalities  (\ref{F-p-1}) and (\ref{F-p-2}) hold for the function $f$ on set (\ref{f-r-s}), they hold for its faithful extension to the set $\mathcal{B}$. This can be easily shown by using the definition of  faithful extension and basic properties of the entropy. So, the assertions of the proposition follow from  Proposition \ref{SCB-1} and Corollary \ref{SCB-G-1}.
\rule{5pt}{5pt}\smallskip

By applying Proposition \ref{LCME} to the entropy  and to the conditional entropy
we obtain the following continuity bounds
\begin{equation}\label{CB-1}
 |H(\rho_A)-H(\sigma_A)|\leq \sqrt{2\varepsilon}F_{H_A}\!\!\left(E/\varepsilon\right)+g(\sqrt{2\varepsilon})
\end{equation}
and
\begin{equation}\label{CB-2}
 |H_\mathrm{e}(A|B)_{\rho}-H_\mathrm{e}(A|B)_{\sigma}|\leq 2\sqrt{2\varepsilon}F_{H_A}\!\!\left(E/\varepsilon\right)+g(\sqrt{2\varepsilon})
\end{equation}
under the conditions $\Tr H_A\rho_A,\Tr H_A\sigma_A\leq E$ and $\,\varepsilon=\frac{1}{2}\|\shs\rho-\sigma\|_1\leq\frac{1}{2}$, where $H_\mathrm{e}(A|B)$ is the faithful extension of the conditional entropy to the set\break
$\,\left\{\shs\rho_{AB}\shs|\,H(\rho_{A})<+\infty\shs\right\}$ defined in (\ref{ce-ext}). These continuity bounds give more rough estimates for variations than the asymptotically tight  continuity bounds for these quantities obtained by Winter in \cite{W-CB}. This is not surprising, since Winter's method does not use  purifications of initial states leading to appearance of the factor $\,\sqrt{\varepsilon}\,$ in (\ref{CB-1}) and (\ref{CB-2}).

The main advantage of Proposition \ref{LCME} is its universality. It allows to obtain continuity bounds \emph{under different forms of energy constrains}. For example, by considering the mutual information $I(A\!:\!B)$ as a function on the set $\S(\H_{ABC})$ and by using the inequality
$0\leq I(A\!:\!B)\leq I(A\!:\!BC)$, upper bound (\ref{MI-UB}) and inequality (\ref{F-c-b})  we obtain from  Proposition \ref{LCME} the following
\smallskip

\begin{corollary}\label{SCB-c} \emph{Let  $ABC$ be a tripartite quantum system and $\,H_{BC}$ a positive operator in $\H_{BC}$ satisfying  condition (\ref{H-cond+}). Then the function
$\rho_{ABC}\mapsto I(A\!:\!B)_{\rho}$ is uniformly continuous on the set $\,\{\shs\rho_{ABC}\,|\,\Tr H_{BC}\rho_{BC}\leq E\shs\}$ for any $\,E\geq E_0\doteq\inf\limits_{\|\varphi\|=1}\langle\varphi|H_{BC}|\varphi\rangle$.  Quantitatively,
\begin{equation}\label{CB-3}
 |I(A\!:\!B)_{\rho}-I(A\!:\!B)_{\sigma}|\leq 2\sqrt{2\varepsilon} F_{H_{BC}}\!\left(E/\varepsilon\right)+2g(\sqrt{2\varepsilon})
\end{equation}
for any states $\rho$ and $\sigma$ in $\S(\H_{ABC})$ such that $\,\Tr H_{BC}\rho_{BC},\Tr H_{BC}\sigma_{BC}\leq E$ and\break $\;\frac{1}{2}\|\shs\rho-\sigma\|_1\leq\varepsilon\leq\frac{1}{2}$, where $F_{H_{BC}}(E)\doteq \!\sup\limits_{\Tr H_{BC}\rho\leq E}\!H(\rho)$.}

\emph{For pure states $\rho$ and $\sigma$ inequality (\ref{CB-3}) holds with $\,\varepsilon$ replaced by $\,\varepsilon^2/2$.}\smallskip
\end{corollary}

By using the Stinespring representation of a quantum channel one can obtain from Corollary \ref{SCB-c}  continuity bound for the output mutual information of a channel under the input energy constraint \emph{not depending on a channel} (see Proposition \ref{MI-CB} in Section 5.3).

\subsection{Relative entropy distances}

\subsubsection{General case}

The relative entropy distance from a  state $\rho$ in $\S(\H)$ to a given subset $\A\subset\S(\H)$ is defined as follows
\begin{equation}\label{red-def}
  D_{\A}(\rho)=\inf_{\omega\in\A}H(\rho\shs\|\shs\omega).
\end{equation}
This function is widely used in quantum information theory for construction of different characteristics of  quantum states \cite{P&V,V&P,SW-1,SW-2}. The most known example is the relative entropy of entanglement of a bipartite state considered in the next subsection.

It is known (cf.\cite{W-CB}) that for any set $\A$ the function $D_{\A}$ satisfies the inequality
\begin{equation}\label{RED-ineq}
D_{\A}(p\rho+(1-p)\sigma)\geq p D_{\A}(\rho)+(1-p)D_{\A}(\sigma)-h_2(p)
\end{equation}
valid for any states $\rho$ and $\sigma$ in $\S(\H)$ and $p\in(0,1)$ with possible values $+\infty$ in both sides. It follows directly from the
analogous inequality for the function $\rho\mapsto H(\rho\shs\|\shs\omega)$ (proved in Lemma \ref{RE-LAA} below in the infinite-dimensional settings) and the
definition (\ref{red-def}) of the function $D_{\A}$.

If the set $\A$ is convex then the joint convexity of the relative entropy implies
convexity of the function $D_{\A}$. So, in this case the function $D_{\A}$ satisfies the LAA-property (\ref{a-a-f}) with $a(p)=h_2(p)$ and $b(p)=0$.
Hence, we obtain from Theorem \ref{AFM} the following infinite-dimensional version of Lemma 7 in \cite{W-CB}.\smallskip

\begin{proposition}\label{RED-UC} \emph{Let $\,H$ be a positive operator in $\H$, $\,\C_{H,E}$  the subset of $\,\S(\H)$ determined by the inequality $\,\Tr H\rho\leq E$, $\,E\geq E_0\doteq\inf\limits_{\|\varphi\|=1}\langle\varphi|H|\varphi\rangle$, and $\,\A$ a convex subset of $\,\S(\H)$. If
$$
G_{H,\A}(E)\doteq \sup_{\Tr H\rho\leq E}D_{\A}(\rho)=o\shs(\sqrt{E})\quad\textit{as}\quad E\rightarrow +\infty\vspace{-5pt}
$$
then the function $\shs D_{\A}$ is uniformly continuous on the set $\,\C_{H,E}$ for any $\,E\geq E_0$ and
\begin{equation}\label{RED-CB}
|D_{\A}(\rho)-D_{\A}(\sigma)|\leq \sqrt{2\varepsilon}G_{H,\A}\!\left(E/\varepsilon\right)+g(\sqrt{2\varepsilon})
\end{equation}
for any states $\rho$ and $\sigma$ in $\,\C_{H,E}$ such that $\,\frac{1}{2}\|\rho-\sigma\|_1\leq\varepsilon\leq \frac{1}{2}$.}\smallskip

\emph{For pure states $\rho$ and $\sigma$ inequality (\ref{RED-CB}) holds with $\shs\varepsilon$ replaced by $\,\varepsilon^2/2$.}
\end{proposition}\smallskip

\textbf{Example}. Let $\mathcal{G}_H$ be the Gibbs family corresponding to a positive operator $H$ satisfying condition (\ref{H-cond+}), i.e.
$\mathcal{G}_H=\left\{\gamma_{H,\lambda}\doteq e^{-\lambda H}/\Tr e^{-\lambda H}\right\}_{\lambda>0}$.
By using   Proposition 1 in \cite{EC} it is easy to show that
\begin{equation}\label{red-gf}
D_{\mathcal{G}_H}(\rho)=H(\rho\shs\|\shs \gamma_{H,\lambda(\rho)})=F_H(\Tr H\rho)-H(\rho),
\end{equation}
for any state $\rho$ with finite "energy" $\Tr H\rho$, where $\gamma_{H,\lambda(\rho)}$ is the Gibbs state such that $\,\Tr H\gamma_{H,\lambda(\rho)}=\Tr H\rho\,$ and $\,F_H(E)\doteq\sup\limits_{\Tr H\rho\leq E}H(\rho)$.
Since the function $\rho\mapsto\Tr H\rho$ is not continuous on $\C_{H,E}$ for any $E>E_0$ (this can be shown by exploiting the sequence $\{\sigma_n\}$ used at the end of the proof of Proposition 1 in \cite{EC}), while the entropy is continuous on $\C_{H,E}$ due to condition (\ref{H-cond+}), the function $D_{\mathcal{G}_H}$ is not continuous on $\C_{H,E}$ for any $E>E_0$.\footnote{The relative entropy distance to Gibbs families may be discontinuous even in the finite-dimensional case \cite{SW-1,SW-2}.}\smallskip

Let $\A$ be any convex set containing the Gibbs family $\mathcal{G}_H$, in particular $\A=\mathrm{conv}(\mathcal{G}_H)$. It follows from (\ref{red-gf})
that
$
D_{\A}(\rho)\leq D_{\mathcal{G}_H}(\rho)\leq F_H(\Tr H\rho).
$
Since condition (\ref{H-cond+}) implies $F_H(E)=o\shs(\sqrt{E})$ as $E\rightarrow +\infty$, Proposition \ref{RED-UC} shows that
the function $D_{\A}$ is uniformly continuous on the set $\,\C_{H,E}$ for any $E\geq E_0$ and
\begin{equation*}
|D_{\A}(\rho)-D_{\A}(\sigma)|\leq \sqrt{2\varepsilon}F_{H}\!\left(E/\varepsilon\right)+g(\sqrt{2\varepsilon})
\end{equation*}
for any states $\rho$ and $\sigma$ in $\,\C_{H,E}$ such that $\,\frac{1}{2}\|\rho-\sigma\|_1\leq\varepsilon\leq\frac{1}{2}$.\footnote{To prove uniform continuity of the function $D_{\A}$ on the set $\,\C_{H,E}$ it suffices to assume that the set $\A$ contains a sequence $\{\gamma_{H,\lambda_n}\}$, in which $\lambda_n$ tends to zero as $\,n\rightarrow\infty$.}\smallskip

\begin{lemma}\label{RE-LAA} \emph{Let $\H$ be a separable Hilbert space and $\omega$ a state in $\S(\H)$. Then
\begin{equation}\label{RE-ineq}
H(p\rho+(1-p)\sigma\shs\|\shs\omega)\geq p H(\rho\shs\|\shs\omega)+(1-p)H(\sigma\shs\|\shs\omega)-h_2(p)
\end{equation}
for any states $\rho$ and $\sigma$ in $\S(\H)$ and $\,p\in(0,1)$ with possible values $+\infty$ in both sides.}
\end{lemma}

\emph{Proof.} If either $\,\supp\shs\rho\,$ or $\,\supp\shs\sigma\,$ is not contained in
$\,\supp\shs\omega\,$ then both sides of (\ref{RE-ineq}) equal to $+\infty$. So, we may assume that $\omega$ is a full rank state.

If $\rho$ and $\sigma$ are finite rank states such that $\,\Tr\sigma\log\omega\,$ and  $\,\Tr\rho\log\omega\,$ are finite then (\ref{RE-ineq}) follows directly from the inequality (\ref{w-k-ineq}), since
in this case we have (cf.\cite{W-CB})
$$
\begin{array}{ccc}
H(p\rho+(1-p)\sigma\shs\|\shs\omega)=-H(p\rho+(1-p)\sigma)-p\Tr\rho\log\omega-(1-p)\Tr\sigma\log\omega\\\\
=p H(\rho\shs\|\shs\omega)+(1-p)H(\sigma\shs\|\shs\omega)+pH(\rho)+(1-p)H(\sigma)-H(p\rho+(1-p)\sigma).
\end{array}
$$
If either $\,\Tr\sigma\log\omega=-\infty\,$ or $\,\Tr\rho\log\omega=-\infty\,$  then both sides of (\ref{RE-ineq}) are equal to $+\infty$. So, (\ref{RE-ineq}) holds for any finite rank states
$\rho$ and $\sigma$.

Let $\rho$ and $\sigma$ be arbitrary states and $\{P_n\}$ a sequence of finite rank projectors strongly converging to the unit operator $I_{\H}$.
Let
$$
\rho_n=a^{-1}_nP_n\rho P_n,\quad \sigma_n=b^{-1}_nP_n\sigma P_n,\quad \omega_n=c^{-1}_nP_n\omega P_n,
$$
and $\,p_n=pa_n/(pa_n+(1-p)b_n)$, where $a_n=\Tr P_n\rho$, $b_n=\Tr P_n\sigma$  and $c_n=\Tr P_n\omega$. For each $n$ by the above observation we have
\begin{equation}\label{RE-ineq-n}
H(p_n\rho_n+(1-p_n)\sigma_n\shs\|\shs\omega_n)\geq p_n H(\rho_n\shs\|\shs\omega_n)+(1-p_n)H(\sigma_n\shs\|\shs\omega_n)-h_2(p_n).
\end{equation}
Since $p_n\rho_n+(1-p_n)\sigma_n=(pa_n+(1-p)b_n)^{-1}P_n (p\rho+(1-p)\sigma) P_n$, by using the lower semicontinuity of the relative entropy and its monotonicity under the map $P_n(\cdot)P_n$ it is easy to show that
$$
\lim_{n\rightarrow\infty}H(\rho_n\shs\|\shs\omega_n)=H(\rho\shs\|\shs\omega),\quad
\lim_{n\rightarrow\infty}H(\sigma_n\shs\|\shs\omega_n)=H(\sigma\shs\|\shs\omega)
$$
and
$$
\lim_{n\rightarrow\infty}H(p_n\rho_n+(1-p_n)\sigma_n\,\|\,\omega_n)=H(p\rho+(1-p)\sigma\,\|\,\omega).
$$
By passing to the limit in (\ref{RE-ineq-n}) we obtain (\ref{RE-ineq}). \rule{5pt}{5pt}

\subsubsection{The relative entropy of entanglement and its regularization}

The relative entropy of entanglement is a one of the main entanglement measures in finite-dimensional bipartite systems. It is defined as follows
\begin{equation}\label{ree-def}
  E_R(\rho)=\inf_{\omega\in \mathcal{S}}H(\rho\shs\|\shs\omega),
\end{equation}
where $\mathcal{S}$ is the set of separable (nonentangled) states in $\S(\H_{AB})$ defined as the convex hull of all product states $\rho_A\otimes\sigma_B$ \cite{P&V,V&P,D,4H}.

The relative entropy of entanglement possesses basic properties of entanglement measures (convexity, LOCC-monotonicity, asymptotic continuity, etc.) but it is nonadditive. The regularization of $E_R$ is defined by the standard way:
\begin{equation}\label{ree-r-def}
  E^{\infty}_R(\rho)=\lim_{n\rightarrow+\infty}n^{-1}E_R(\rho^{\otimes n}).
\end{equation}

Fannes' type continuity bounds for $E_R(\rho)$ and $E^{\infty}_R(\rho)$ have been obtained in \cite{D}. Recently Winter essentially refined these continuity bounds by using the AFW-method \cite{W-CB}. He proved that
$$
|E(\rho)-E(\sigma)|\leq\varepsilon \log d+g(\varepsilon),\quad E=E_R,E^{\infty}_R,
$$
for any states $\rho$ and $\sigma$, where $\,d=\min\{\dim\H_A,\dim\H_B\}$ and $\varepsilon=\frac{1}{2}\|\shs\rho-\sigma\|_1$.

Definitions (\ref{ree-def}) and (\ref{ree-r-def}) are valid in the case $\,\dim\H_A=\dim\H_B=+\infty$. One should only to note that in this case the set $\mathcal{S}$ of separable states is defined as the convex closure of all product states in $\S(\H_{AB})$. The above mentioned Winter's result shows that
$E_R$ and $E^{\infty}_R$ are uniformly continuous on the set $\S(\H_{AB})$ if (and only if) one of the systems, say system $A$, is finite dimensional.
It is also known that $E_R$ is continuous on the set of states with bounded energy of $\rho_A$ and of $\rho_B$ provided the Hamiltonians of both subsystems satisfies condition (\ref{H-cond}) \cite{E&Co}.
By using the modification of the AFW-method one can substantially strengthen the above results.\smallskip

\begin{proposition}\label{REoE} \emph{Let $A$ and $B$ be infinite-dimensional quantum systems, $H_A$ the Hamiltonian of system $A$ satisfying condition (\ref{H-cond+}) and $E_0\doteq\inf\limits_{\|\varphi\|=1}\langle\varphi|H_A|\varphi\rangle$. Then the functions  $E_R$ and $E^{\infty}_R$ (defined respectively in (\ref{ree-def}) and (\ref{ree-r-def})) are uniformly continuous on the set $\,\{\shs\rho_{AB}\shs|\,\Tr H_A\rho_{A}\leq E\shs\}$ for any $E\geq E_0$.  Quantitatively,
\begin{equation}\label{RE-CB}
|E(\rho)-E(\sigma)|\leq\sqrt{2\varepsilon} F_{H_A}\!\!\left(E/\varepsilon\right)+g(\sqrt{2\varepsilon}),\quad E=E_R,E^{\infty}_R,
\end{equation}
for any states $\shs\rho\shs$ and $\shs\sigma$  such that $\,\Tr H_A\rho_{A},\Tr H_A\sigma_{A}\leq E\,$ and $\;\frac{1}{2}\|\shs\rho-\sigma\|_1\leq\varepsilon\leq\frac{1}{2}$, where $\,F_{H_A}(E)=\sup\limits_{\Tr H_{A}\rho\leq E}H(\rho)\,$  and  $\,g(x)\!=\!(1+x)h_2\!\left(\frac{x}{1+x}\right)$.}

\emph{If $\,A$ is the $\ell$-mode quantum oscillator then the function $F_{H_A}$ in (\ref{RE-CB}) can be replaced by its upper bound $\widehat{F}_{\ell,\omega}$ defined in (\ref{bF-ub})}.\smallskip

\end{proposition}

\emph{Proof.} All the assertions for $E=E_R$ directly follow from Proposition \ref{SCB-1}, since the inequality
\begin{equation}\label{RE-UB}
0\leq E_R(\rho_{AB})\leq H(\rho_A)
\end{equation}
(see \cite{P&V,V&P}) along with the convexity of $E_R$ and Lemma \ref{RE-LAA} show that the function $f=E_R$ satisfies (\ref{F-p-1}) and (\ref{F-p-2}) with $a_f=1$, $b_f=0$, $c^-_f=0$, $c^+_f=1$ and $B=A$.

To prove continuity bound (\ref{RE-CB}) for $E=E^{\infty}_R$ we will use the telescopic method from the proof of Corollary 8 in \cite{W-CB} with necessary modifications. For given natural $n$ we have
$$
\!\!\begin{array}{c}
E_R(\rho^{\otimes n})-E_R(\sigma^{\otimes n})\leq\displaystyle\sum_{k=1}^n \left|E_R\left(\rho^{\otimes k}\otimes\sigma^{\otimes(n-k)}\right)-E_R\left(\rho^{\otimes (k-1)}\otimes\sigma^{\otimes(n-k+1)}\right)\right|\\\\\displaystyle\leq\sum_{k=1}^n \left|E_R\left(\rho\otimes\omega_k\right)-E_R\left(\sigma\otimes\omega_k\right)\right|,
\end{array}
$$
where $\omega_k=\rho^{\otimes (k-1)}\otimes\sigma^{\otimes(n-k)}$. The assumption
$\,\Tr H_A\rho_{A},\Tr H_A\sigma_{A}\leq E$ and inequality (\ref{RE-UB}) imply finiteness of all the terms in the above inequality.
So, to prove the continuity bound for $E^{\infty}_R$ it suffices to show that
\begin{equation}\label{RE-T}
\left|E_R\left(\rho\otimes\omega_k\right)-E_R\left(\sigma\otimes\omega_k\right)\right|\leq\sqrt{2\varepsilon} F_{H_A}\!\left(E/\varepsilon\right)+g(\sqrt{2\varepsilon})
\end{equation}
for each $k$. This can be made by repeating the arguments from the proof of Theorem 1.

Let  $\hat{\rho}$ and $\hat{\sigma}$ be purifications of the states $\rho$ and $\sigma$ such that  $\delta\doteq\frac{1}{2}\|\hat{\rho}-\hat{\sigma}\|_1=\!\sqrt{2\varepsilon}$.
Then $\hat{\varrho}_k=\hat{\rho}\otimes\hat{\omega}_k$ and $\hat{\varsigma}_k=\hat{\sigma}\otimes\hat{\omega}_k$, where $\hat{\omega}_k=\hat{\rho}^{\otimes (k-1)}\otimes\hat{\sigma}^{\otimes(n-k)}$, are
purifications of the states $\varrho_k\doteq\rho\otimes\omega_k$ and $\varsigma_k\doteq\sigma\otimes\omega_k$ such that $\frac{1}{2}\|\hat{\varrho}_k-\hat{\varsigma}_k\|_1=\delta$.

Let $\hat{\tau}_{\pm}=\delta^{-1}[\shs\hat{\rho}-\hat{\sigma}\shs]_{\pm}$ and $\tau_{\pm}=[\hat{\tau}_{\pm}]_{AB}$.  The estimation in the proof of Theorem 1 shows that $\Tr H_A[\tau_{\pm}]_A\leq E/\varepsilon$. Hence inequality (\ref{RE-UB}) implies
\begin{equation}\label{RE-T-3}
E_R(\tau_{\pm})\leq H([\tau_{\pm}]_A)\leq F_{H_A}(E/\varepsilon)<+\infty.
\end{equation}

By applying the main trick from the proof of Theorem 1 to the states $\hat{\varrho}_k$, $\hat{\varsigma}_k$ and
$\delta^{-1}[\shs\hat{\varrho}_k-\hat{\varsigma}_k\shs]_{\pm}=\hat{\tau}_{\pm}\otimes\hat{\omega}_k$ (instead of $\hat{\rho}$, $\hat{\sigma}$ and
$\hat{\tau}_{\pm}$) and by using the  convexity of $E_R$ and inequality (\ref{RED-ineq}) with $D_{\C}=E_R$ we obtain
\begin{equation}\label{RE-T-1}
\left|E_R(\varrho_k)-E_R(\varsigma_k)\right|\leq \delta\left|E_R(\tau_+\!\otimes\omega_k)-E_R(\tau_-\!\otimes\omega_k)\right|+g(\delta).
\end{equation}
Assume that $E_R(\tau_+\!\otimes\omega_k)\geq E_R(\tau_-\!\otimes\omega_k)$. Then the subadditivity of $E_R$ implies that
$\,E_R(\tau_+\!\otimes\omega_k)\leq E_R(\tau_+)+E_R(\omega_k)$,
while the LOCC-monotonicity of $E_R$ shows that $E_R(\tau_-\!\otimes\omega_k)\geq E_R(\omega_k)$ (cf.\cite{W-CB}). Hence
\begin{equation}\label{RE-T-2}
|E_R(\tau_+\!\otimes\omega_k)-E_R(\tau_-\!\otimes\omega_k)|\leq \max\left\{E_R(\tau_-),E_R(\tau_+)\right\}.
\end{equation}
Inequalities (\ref{RE-T-3}),(\ref{RE-T-1}) and (\ref{RE-T-2}) imply (\ref{RE-T}). \rule{5pt}{5pt}\smallskip

Proposition \ref{REoE} implies the following asymptotic continuity property of the relative entropy of  entanglement and of
its regularization (cf.\cite{E&Co}).\smallskip

\begin{corollary}\label{REoE-ac} \emph{Let $\,\{\rho_{n}\}\,$ and $\,\{\sigma_{n}\}\,$ be any sequences of states such that
$$
\rho_{n},\sigma_{n}\in\S(\H^{\otimes n}_{AB}),\quad \Tr H_{A^n}[\rho_{n}]_{A^n},\Tr H_{A^n}[\sigma_{n}]_{A^n}\leq nE, \quad \lim_{n\rightarrow\infty}\|\rho_{n}-\sigma_{n}\|_1=0,
$$
where $\,H_{A^n}=H_A\otimes I_A\otimes\ldots\otimes I_A+\ldots+I_A\otimes\ldots\otimes I_A\otimes H_A$
is the Hamiltonian of the system $A^{n}$. If $\,H_A$ satisfies condition (\ref{H-cond+}) then
$$
\lim_{n\rightarrow\infty}\frac{\left|E_R(\rho_{n})-E_R(\sigma_{n})\right|}{n}=0\quad \textit{and}\quad
\lim_{n\rightarrow\infty}\frac{\left|E_R^{\infty}(\rho_{n})-E_R^{\infty}(\sigma_{n})\right|}{n}=0.
$$
In particular, these relations hold if $\,A$ is the $\ell$-mode quantum oscillator.}
\end{corollary}\smallskip

\emph{Proof.} Note that
$$
F_{H_{A^n}}(nE)=H(\gamma_{A^n}(nE))=H([\gamma_A(E)]^{\otimes n})=nH(\gamma_A(E))=nF_{H_{A}}(E).
$$
Since $\,H_A$ satisfies condition (\ref{H-cond+}), we have $F_{H_{A}}(E)=o\shs(\sqrt{E})$ as $E\rightarrow\infty$ (by Lemma \ref{SCB-l-1}).  So, the required limit relations  follow directly from the continuity bounds in Proposition \ref{REoE}. \rule{5pt}{5pt}

\subsection{Continuity bound for the mutual information at the output of a channel}

A \emph{quantum channel}  from a system $A$ to a system
$B$ is a completely positive trace preserving linear map from
$\mathfrak{T}(\mathcal{H}_A)$ into $\mathfrak{T}(\mathcal{H}_B)$ \cite{H-SCI,N&Ch,Wilde}.
In analysis of information properties of a  channel $\,\Phi:A\rightarrow B\,$ the quantity $I(B\!:\!C)_{\Phi\otimes\mathrm{Id}_{R}(\rho)}$
is widely used, where $C$ is a given quantum system and $\rho$ is a state in $\S(\H_{AR})$ \cite{H-SCI,Wilde}.

If $\Phi$ is an infinite-dimensional quantum channel (i.e. $\dim\H_A=\dim\H_B=+\infty$) then the function $\rho\mapsto I(B\!:\!C)_{\Phi\otimes\mathrm{Id}_{R}(\rho)}$
is typically considered on the set of states with bounded energy of $A$, i.e. states $\rho$ satisfying the inequality
\begin{equation}\label{input-b-e}
\Tr H_A\rho_A\leq E,
\end{equation}
where $H_A$ is the Hamiltonian of the input system $A$. By using Winter's continuity bound for the conditional entropy under the energy constraint obtained in \cite{W-CB} it is easy to write continuity bound for the function $\rho\mapsto I(B\!:\!C)_{\Phi\otimes\mathrm{Id}_{C}(\rho)}$ under the constraint (\ref{input-b-e}) provided that
\begin{equation}\label{faf}
\sup_{\Tr H_A\rho_A\leq E}\Tr H_B\Phi(\rho_A)<+\infty
\end{equation}
where $H_B$ is the Hamiltonian of the output system $B$.

In this section we show that the modified AFW-method gives continuity bound
for the function $\rho\mapsto I(B\!:\!R)_{\Phi\otimes\mathrm{Id}_{C}(\rho)}$ under the constraint (\ref{input-b-e})
valid for \emph{arbitrary} channel $\Phi$ (and not depending on channel $\Phi$ at all) provided that the Hamiltonian $H_A$ satisfies condition (\ref{H-cond+}).

For any  quantum channel $\,\Phi:A\rightarrow B\,$ the Stinespring theorem implies existence of a Hilbert space
$\mathcal{H}_E$ and of an isometry
$V:\mathcal{H}_A\rightarrow\mathcal{H}_B\otimes\mathcal{H}_E$ such
that
\begin{equation*}
\Phi(\rho)=\mathrm{Tr}_{E}V\rho V^{*},\quad
\rho\in\mathfrak{T}(\mathcal{H}_A).
\end{equation*}

By using this representation and identifying the space $\H_A$ with the subspace $V\H_A$ of $\H_{BE}$ it is easy to derive from  Corollary \ref{SCB-c} in Section 5.1  the following\smallskip

\begin{proposition}\label{MI-CB} \emph{Let $\,\Phi:A\rightarrow B$ be an arbitrary quantum channel and $\,C$ be any system.
If the Hamiltonian $H_{A}$ of input system $A$ satisfies condition (\ref{H-cond+}) then the function
$\,\rho_{AC}\mapsto I(B\!:\!C)_{\Phi\otimes\id_{C}(\rho)}$ is uniformly continuous on the set of states with bounded energy of $\,\rho_{A}$.  Quantitatively,
\begin{equation}\label{MI-CB+}
|I(B\!:\!C)_{\Phi\otimes\id_{C}(\rho)}-I(B\!:\!C)_{\Phi\otimes\id_{C}(\sigma)}|\leq 2\sqrt{2\varepsilon} F_{H_A}\!\!\left(E/\varepsilon\right)+2g(\sqrt{2\varepsilon})\!\!
\end{equation}
for any states $\rho$ and $\sigma$ in $\S(\H_{AC})$ such that $\Tr H_{A}\rho_{A},\Tr H_{A}\sigma_{A}\leq E$ and\break   $\;\frac{1}{2}\|\shs\rho-\sigma\|_1\leq\varepsilon\leq\frac{1}{2}$, where $F_{H_A}(E)=\sup\limits_{\Tr H_A\rho\leq E}H(\rho)$.}

\emph{For pure states $\rho$ and $\sigma$ inequality (\ref{MI-CB+}) holds with $\shs\varepsilon$ replaced by $\,\varepsilon^2/2$.}
\end{proposition}\smallskip

Since the Hamiltonian $H_{A}$ satisfies condition (\ref{H-cond+}), Lemma \ref{SCB-l-1} implies that the main term in (\ref{MI-CB+}) tends to zero as $\,\varepsilon\!\rightarrow\!0$.

By the Bennett-Shor-Smolin-Thaplyal theorem (cf. \cite{BSST}) the entanglement-assisted classical capacity of a quantum channel $\,\Phi:A\rightarrow B\,$ is expressed via the quantum mutual information of this channel at a state $\rho\in\S(\H_A)$ defined as follows
\begin{equation}\label{mi-def}
 I(\Phi,\rho)=I(B\!:\!R)_{\Phi\otimes\mathrm{Id}_{R}(\hat{\rho})},
\end{equation}
where $\mathcal{H}_R\cong\mathcal{H}_A$ and $\hat{\rho}\shs$ is a pure state in $\S(\H_{AR})$ such that $\hat{\rho}_{A}=\rho$. This quantity is well defined and finite for any infinite-dimensional channel $\Phi$ and any input state $\rho$ with finite entropy. So, it can be also used to express the coherent information of $\Phi$ at any such $\rho$ by the formula $I(\Phi,\rho)-H(\rho)$ \cite{H-SCI,CMI}.

Since for any states $\rho$ and $\sigma$ in $\S(\H_{A})$ such that $\,\frac{1}{2}\|\shs\rho-\sigma\|_1\leq\varepsilon\,$ one can find purifications
$\hat{\rho}$ and $\hat{\sigma}$ in $\S(\H_{AR})$ such that $\,\frac{1}{2}\|\shs\hat{\rho}-\hat{\sigma}\|_1\leq\sqrt{2\varepsilon}$,
the last assertion of Proposition \ref{MI-CB} implies the following continuity bound for the function $\rho\mapsto I(\Phi,\rho)$.\smallskip

\begin{corollary}\label{CMI-CB} \emph{Let $\,\Phi:A\rightarrow B$ be an arbitrary quantum channel.
If the Hamiltonian $H_{A}$ of input system $A$ satisfies condition (\ref{H-cond+}) then the function
$\,\rho\mapsto I(\Phi,\rho)$ is uniformly continuous on the set of input states with bounded energy.  Quantitatively,
\begin{equation}\label{CMI-CB+}
|I(\Phi,\rho)-I(\Phi,\sigma)|\leq 2\sqrt{2\varepsilon} F_{H_A}\!\!\left(E/\varepsilon\right)+2g(\sqrt{2\varepsilon})
\end{equation}
for any states $\rho$ and $\sigma$ in $\S(\H_{A})$ such that $\Tr H_{A}\rho\leq E,\Tr H_{A}\sigma\leq E$ and  $\;\frac{1}{2}\|\shs\rho-\sigma\|_1\leq\varepsilon\leq\frac{1}{2}$, where $F_{H_A}(E)=\sup\limits_{\Tr H_A\rho\leq E}H(\rho)$.}

\emph{If $\,A$ is the $\ell$-mode quantum oscillator then the function $F_{H_A}$ in (\ref{CMI-CB+}) can be replaced by its upper bound $\,\widehat{F}_{\ell,\omega}$ defined in (\ref{bF-ub})}.
\end{corollary}\smallskip

It is essential that continuity bounds (\ref{MI-CB+}) and (\ref{CMI-CB+}) \emph{\emph{do not depend on a channel}} $\Phi$.

\subsection{Continuity bound for the output Holevo quantity
not depending on a channel}

A finite or
countable collection $\{\rho_{i}\}$ of quantum states
with a  probability distribution $\{p_{i}\}$ is called \emph{ensemble} and denoted $\{p_i,\rho_i\}$. The state $\bar{\rho}=\sum_{i} p_i\rho_i$ is called  \emph{average state} of  $\{p_i,\rho_i\}$ \cite{H-SCI,Wilde}.\smallskip

Let $\Phi:A\rightarrow B$ be a quantum channel and $\{p_i,\rho_i\}$ an ensemble of states in $\S(\H_A)$. The Holevo quantity\footnote{The Holevo quantity of ensemble of quantum states gives the upper bound for the classical information obtained from  quantum measurements over the ensemble \cite{H-73}.} of the output ensemble
$\{p_i,\Phi(\rho_i)\}$  given by the formula
$$
\chi(\{p_i,\Phi(\rho_i)\})= \sum_{i} p_i H(\Phi(\rho_i)\|\Phi(\bar{\rho}))
$$
plays a basic role in analysis of transmission of classical information through the channel $\Phi$ \cite{H-SCI,Wilde}.

Dealing with infinite-dimensional channels it is natural to consider input ensembles with bounded average energy, i.e. such ensembles $\{p_i,\rho_i\}$ that
\begin{equation}\label{input-b-ae}
\sum_ip_i\Tr H_A\rho_i=\Tr H_A\bar{\rho}\leq E,
\end{equation}
where $H_A$ is the Hamiltonian of the input system $A$.

By using Winter's type  continuity bound for the Holevo quantity under the average energy constraint obtained in \cite{CHI} it is easy to write continuity bound for the function $\{p_i,\rho_i\}\mapsto \chi(\{p_i,\Phi(\rho_i)\})$ under the constraint (\ref{input-b-ae}) provided that the channel $\Phi$ satisfies condition (\ref{faf}). In this section we show that by using Proposition \ref{MI-CB} one can obtain continuity bound for the function\break $\{p_i,\rho_i\}\mapsto \chi(\{p_i,\Phi(\rho_i)\})$ under the constraint (\ref{input-b-ae}) valid for \emph{arbitrary} channel $\Phi$ (and not depending on $\Phi$).

We will use two measures of divergence between  ensembles $\mu=\{p_i,\rho_i\}$ and $\nu=\{q_i,\sigma_i\}$.
The quantity
\begin{equation*}
D_0(\mu,\nu)\doteq\frac{1}{2}\sum_i\|\shs p_i\rho_i-q_i\sigma_i\|_1
\end{equation*}
is a true metric on the set of all  ensembles of quantum states considered as \emph{ordered} collections of states with the corresponding probability distributions. It coincides (up to the factor $1/2$) with the trace norm of the difference between the corresponding $qc$-states $\sum_{i} p_i\rho_i\otimes |i\rangle\langle i|$
and $\sum_{i} q_i\sigma_i\otimes |i\rangle\langle i|$ \cite{Wilde}.\smallskip

The main advantage of $D_0$ is a direct computability, but from the quantum information point of view we have to consider an ensemble of quantum states $\{p_i,\rho_i\}$ as a discrete probability measure $\sum_i p_i\delta(\rho_i)$  on the set $\S(\H)$ (where $\delta(\rho)$ is the Dirac measure concentrating at a state $\rho$) rather than ordered (or disordered) collection of states. If we want to identify ensembles corresponding to the same probability measure then it is natural to use the factorization of $D_0$, i.e. the quantity
 \begin{equation}\label{f-metric}
D_*(\mu,\nu)\doteq \inf_{\mu'\in \mathcal{E}(\mu),\,\nu'\in \mathcal{E}(\nu)}D_0(\mu',\nu')
\end{equation}
as a measure of divergence between ensembles $\mu=\{p_i,\rho_i\}$ and $\nu=\{q_i,\sigma_i\}$, where $\mathcal{E}(\mu)$ and $\mathcal{E}(\nu)$ are the sets
of all countable ensembles corresponding to the measures $\sum_i p_i\delta(\rho_i)$ and $\sum_i q_i\delta(\sigma_i)$ respectively.

It is mentioned in \cite{CHI} that the factor-metric $D_*$ coincides with the EHS\nobreakdash-\hspace{0pt}distance $D_{\mathrm{ehs}}$ between ensembles of quantum states proposed by Oreshkov and Calsamiglia in \cite{O&C}. By using this coincidence and other results from \cite{O&C} it is  shown in \cite{CHI} that $D_*$ generates the weak convergence topology on the set of all ensembles (considered as probability measures).\footnote{This means that a  sequence $\{\{p^n_i,\rho^n_i\}\}_n$  converges to an ensemble $\{p^0_i,\rho^0_i\}$ with respect to the metric $\,D_*\,$ if and only if
$\,\lim_{n\rightarrow\infty}\sum_i p^n_if(\rho^n_i)=\sum_i p^0_if(\rho^0_i)\,$
for any continuous bounded function $f$ on $\,\S(\H)$.}

The metric $\,D_*=D_{\mathrm{ehs}}\,$ is more adequate for continuity analysis of the Holevo quantity, but difficult to compute in general.\footnote{For finite ensembles it can be calculated by a linear programming procedure \cite{O&C}.} It is clear that
\begin{equation}\label{d-ineq}
D_*(\mu,\nu)\leq D_0(\mu,\nu)
\end{equation}
for any ensembles $\mu$ and $\nu$. But in some cases the metrics $\,D_0\,$ and $\,D_*\,$  are close to each other or even coincide. This holds, for example,  if we consider small perturbations  of states or probabilities of a given ensemble.

In the following corollary  we assume that the set of all  ensembles is equipped with
the weak convergence topology  generated by the metric $\,D_*$.\smallskip\pagebreak

\begin{corollary}\label{HQ-CB} \emph{Let $\,\Phi:A\rightarrow B$ be a quantum channel.
If the Hamiltonian $H_{A}$ of the input system $A$ satisfies condition (\ref{H-cond+}) then the function
$\{p_i,\rho_i\}\rightarrow \chi(\{p_i,\Phi(\rho_i)\})$ is uniformly continuous on the set of all ensembles
$\{p_i,\rho_i\}$ with bounded average  energy $\,E(\{p_i,\rho_i\})\doteq\sum_i p_i\Tr H_A\rho_i$. Quantitatively,
\begin{equation}\label{HQ-CB+}
\left|\chi(\{p_i,\Phi(\rho_i)\})-\chi(\{q_i,\Phi(\sigma_i)\})\right|\leq 2\sqrt{2\varepsilon}F_{H_A}\!\!\left(E/\varepsilon\right)+2g(\sqrt{2\varepsilon})
\end{equation}
for any input  ensembles $\{p_i,\rho_i\}$ and $\{q_i,\sigma_i\}$ such that $E(\{p_i,\rho_i\}),E(\{q_i,\sigma_i\})\leq E$ and    $\;D_*(\{p_i,\rho_i\},\{q_i,\sigma_i\})\leq\varepsilon\leq\frac{1}{2}$, where $F_{H_A}(E)=\sup\limits_{\Tr H_A\rho\leq E}H(\rho)$.}

\emph{If $\,A$ is the $\ell$-mode quantum oscillator then the function $F_{H_A}$ in (\ref{HQ-CB+}) can be replaced by its upper bound $\widehat{F}_{\ell,\omega}$ defined in (\ref{bF-ub})}.\smallskip

\emph{The metric $D_*$ in (\ref{HQ-CB+}) can be replaced by the metric $D_0$.}\smallskip
\end{corollary}

\emph{Proof.} Since the Hamiltonian $H_{A}$ satisfies condition (\ref{H-cond+}), Lemma \ref{SCB-l-1} shows that  $\,\sqrt{\varepsilon} F_{H_A}\!\left(E/\varepsilon\right)\rightarrow0\,$ as $\,\varepsilon\rightarrow0$.
So,  continuity bound (\ref{HQ-CB+})  implies uniform continuity of the function
$\{p_i,\rho_i\}\rightarrow \chi(\{p_i,\Phi(\rho_i)\})$ on the set of all ensembles
with bounded average energy.

Take arbitrary
$\epsilon>0$. Let  $\,\{\tilde{p}_i,\tilde{\rho}_i\}$ and $\{\tilde{q}_i,\tilde{\sigma}_i\}$ be ensembles belonging respectively to the sets $\mathcal{E}(\{p_i,\rho_i\})$ and $\mathcal{E}(\{q_i,\sigma_i\})$ such that
$D_*(\{p_i,\rho_i\}, \{q_i,\sigma_i\})\geq D_0(\{\tilde{p}_i,\tilde{\rho}_i\}, \{\tilde{q}_i,\tilde{\sigma}_i\})-\epsilon$
(see definition (\ref{f-metric}) of $D_*$). Consider the $qc$-states
$$
\hat{\rho}=\sum_{i} \tilde{p}_i\tilde{\rho}_i\otimes |i\rangle\langle i|\quad
\textrm{and}\quad \hat{\sigma}=\sum_{i}\tilde{q}_i\tilde{\sigma}_i\otimes |i\rangle\langle i|
$$
in $\S(\H_{AC})$, where $\{|i\rangle\}$ is a basic in $\H_C$. We have
$$
\chi(\{p_i,\Phi(\rho_i)\})=\chi(\{\tilde{p}_i,\Phi(\tilde{\rho}_i)\})=I(B\!:\!C)_{\Phi\otimes\id_C(\hat{\rho})}
$$
and
$$
\chi(\{q_i,\Phi(\sigma_i)\})=\chi(\{\tilde{q}_i,\Phi(\tilde{\sigma}_i)\})=I(B\!:\!C)_{\Phi\otimes\id_C(\hat{\sigma})}.
$$
Since $\,\|\hat{\rho}-\hat{\sigma}\|_1=2D_0(\{\tilde{p}_i,\tilde{\rho}_i\}, \{\tilde{q}_i,\tilde{\sigma}_i\})$, $E(\{p_i,\rho_i\})=E(\{\tilde{p}_i,\tilde{\rho}_i\})=\Tr H_A\hat{\rho}_A$ and $E(\{q_i,\sigma_i\})=E(\{\tilde{q}_i,\tilde{\sigma}_i\})=\Tr H_A\hat{\sigma}_A$, continuity bound (\ref{HQ-CB+}) follows from continuity bound (\ref{MI-CB+}).

The last assertion of the proposition follows from (\ref{d-ineq}). \rule{5pt}{5pt}

\subsection{On other applications}

The modification of the AFW-method described in Sections 3,4 is a basic tool of the proof of the uniform finite-dimensional approximation theorem for basic capacities of energy-constrained channels presented in \cite{UFA}. This theorem states, briefly speaking, that
dealing with basic capacities of energy-constrained channels we may assume (accepting arbitrarily small error $\varepsilon$) that all channels have the same finite-dimensional input space -- the subspace corresponding to the $m(\varepsilon)$ minimal eigenvalues of the input Hamiltonian (which is assumed to satisfy condition (\ref{H-cond+})).

In particular, this theorem allows to prove the uniform continuity of the basic capacities on the set of \emph{all} quantum channels with respect to the strong (pointwise) convergence topology (see details in \cite{UFA}).

\section*{Appendix}

\textbf{Proof of Lemma \ref{SCB-l-1}.} Show first that condition (\ref{H-cond+}) implies
\begin{equation}\label{F-H-a}
  F_{H_B}(E)\doteq\sup_{\Tr H_B\rho<E}H(\rho)=o(\sqrt{E})\quad \textrm{as}\quad E\rightarrow+\infty.
\end{equation}
Condition (\ref{H-cond+}) shows that $\Tr e^{-\lambda H_B}<+\infty$ for all $\lambda>0$. So, the operator $H_B$ has the discrete spectrum $\{E_k\}_{k\geq0}$. We may assume that $E_{k+1}\geq E_k$ for all $k$. Condition (\ref{H-cond+}) means that
\begin{equation}\label{one}
\lim_{\lambda\rightarrow+0}\lambda g(\lambda)=0,\quad
\textrm{where}\quad g(\lambda)=\log \sum_{k=0}^{+\infty}e^{-\lambda E_k}.
\end{equation}

It is shown in the proof of Proposition 1 in \cite{EC} that
$\,F^{\shs\prime}_{H_B}(E)=\lambda(E)\,$ for all $E$ in $[E_0,+\infty)$, where $\lambda(E)$ is a smooth strictly decreasing function  determined by the equality
\begin{equation}\label{l-eq}
\sum_{k=0}^{+\infty}E_k e^{-\lambda E_k}=E\sum_{k=0}^{+\infty}e^{-\lambda E_k}
\end{equation}
such that
\begin{equation}\label{g-pr}
\lim_{E\rightarrow E_0+0}\lambda(E)=+\infty\quad \textrm{and}\quad \lim_{E\rightarrow +\infty}\lambda(E)=0.
\end{equation}
By L'Hopital's rule to prove that $F_{H_B}(E)=o(\sqrt{E})$ it suffices to show that
\begin{equation}\label{hr}
\lim_{E\rightarrow+\infty}\sqrt{E}\lambda(E)=0.
\end{equation}
Denote by $E(\lambda)$ the inverse function to $\lambda(E)$. Equality  (\ref{l-eq}) implies that
\begin{equation}\label{b-eq}
  E(\lambda)=-g^{\shs\prime}(\lambda),
\end{equation}
where $g(\lambda)$ is the function defined in (\ref{one}). It follows  from  (\ref{g-pr}) and (\ref{b-eq}) that  (\ref{hr}) can be rewritten as
\begin{equation}\label{two}
\lim_{\lambda\rightarrow+0}\lambda^2g^{\shs\prime}(\lambda)=0.
\end{equation}
So, to prove the lemma it suffices to show that (\ref{one}) implies (\ref{two}).
Assume that (\ref{two}) is not valid. Then there exists a vanishing sequence $\{\lambda_n\}$ of positive numbers such that
$\lambda_n^2|g^{\shs\prime}(\lambda_n)|\geq\delta>0$ for all $n$. Since (\ref{b-eq}) and the strict concavity of $F_{H_B}$ imply that
$$
g^{\shs\prime\prime}(\lambda)=-E^{\shs\prime}(\lambda)=-1/\lambda^{\shs\prime}(E)=-1/F_{H_B}^{\shs\prime\prime}(E)>0,
$$
the positive function $g(\lambda)$ is convex. It follows that for any $\lambda_n$ and $\lambda\in(0,\lambda_n)$ we have
$$
g(\lambda)\geq g(\lambda_n)+|g^{\shs\prime}(\lambda_n)|(\lambda_n-\lambda)\geq g(\lambda_n)+\delta(\lambda_n-\lambda)/\lambda^2_n
$$
and hence
$$
\lambda g(\lambda)\geq \lambda g(\lambda_n)+\delta\lambda(\lambda_n-\lambda)/\lambda^2_n\geq \delta\lambda(\lambda_n-\lambda)/\lambda^2_n.
$$
By taking $\lambda=\lambda_n/2$ we obtain $\,(\lambda_n/2)g(\lambda_n/2)\geq\delta/4\,$ for all $n$ contradicting to (\ref{one}).\smallskip

Show that condition (\ref{F-H-a}) implies (\ref{H-cond+}). It follows from (\ref{F-exp}) and (\ref{F-H-a}) that
\begin{equation}\label{T-exp}
\lambda(E)\sqrt{E}+\frac{\log[\Tr e^{-\lambda(E) H_{\!B}}]^{\lambda(E)}}{\lambda(E)\sqrt{E}}\,\rightarrow\,0\quad\textup{as}\quad E\rightarrow+\infty.
\end{equation}
By Proposition 1 in \cite{EC} condition (\ref{F-H-a}) implies (\ref{H-cond}), which guarantees that $\lambda(E)$ is a strictly decreasing smooth function on $[E_0,+\infty)$ vanishing as $E\rightarrow+\infty$. Hence the second summand in (\ref{T-exp}) is nonnegative for large $E$. This implies that $\lambda(E)\sqrt{E}$ tends to zero as $E\rightarrow+\infty$. It follows from  (\ref{T-exp}) that  $\log[\Tr e^{-\lambda(E) H_{\!B}}]^{\lambda(E)}$ also tends to zero as $E\rightarrow+\infty$.  This and the above-mentioned properties of the function $\lambda(E)$ imply (\ref{H-cond+}). \rule{5pt}{5pt}

\smallskip

\textbf{Proof of Lemma \ref{SCB-l-2}.} Note that $\,\sum_{k\geq 1} e^{-\lambda E_k}<+\infty\,$ for all $\,\lambda>0\,$ if  and only if $\,q>1$.

For any $\,q>1$ we have
\begin{equation}\label{int-est}
  \int_{1}^{+\infty}e^{-\lambda \log^q x}dx \leq \sum_{k=1}^{+\infty}e^{-\lambda E_k}\leq \int_{1}^{+\infty}e^{-\lambda \log^q x}dx +1.
\end{equation}

By introducing the variable $\,u=\lambda^{1/q}\log x\,$ we obtain
$$
I(\lambda)\doteq\int_{1}^{+\infty}e^{-\lambda \log^q x}dx=\lambda^{-1/q}\int_{0}^{+\infty}e^{-u^q+u\lambda^{-1/q}}du.
$$

If $\,q>2\,$ then
$$
\int_{0}^{1}e^{-u^q+u\lambda^{-1/q}}du\leq \int_{0}^{1}e^{u\lambda^{-1/q}}du=\lambda^{1/q}[e^{\lambda^{-1/q}}-1]
$$
and
$$
\begin{array}{c}
\displaystyle\int_{1}^{+\infty}e^{-u^q+u\lambda^{-1/q}}du\leq \int_{1}^{+\infty}e^{-u^2+u\lambda^{-1/q}}du\\\\
\displaystyle=\int_{1}^{+\infty}e^{-(u-0.5\lambda^{-1/q})^2+0.25\lambda^{-2/q}}du\leq e^{0.25\lambda^{-2/q}}\int_{-\infty}^{+\infty}e^{-t^2}dt=\sqrt{\pi}e^{0.25\lambda^{-2/q}}.
\end{array}
$$
Since $2/q<1$, these estimates  show that $\,\lim_{\lambda\rightarrow+0}\lambda\log I(\lambda)=0$. Hence the right inequality in (\ref{int-est}) implies
\begin{equation}\label{last-eq}
\lim_{\lambda\rightarrow+0}\left[\sum_k e^{-\lambda E_k}\right]^{\lambda}=1
\end{equation}
 in this case.\smallskip

If $\,q=2\,$ then
$$
\begin{array}{c}
\displaystyle I(\lambda)=\lambda^{-1/2}\int_{0}^{+\infty}e^{-u^2+u\lambda^{-1/2}}du=\lambda^{-1/2}
\int_{0}^{+\infty}e^{-(u-0.5\lambda^{-1/2})^2+0.25\lambda^{-1}}du\\\\\displaystyle\geq \lambda^{-1/2} e^{0.25\lambda^{-1}}\int_{0}^{+\infty}e^{-t^2}dt=\frac{\sqrt{\pi}}{2}\,\lambda^{-1/2}e^{0.25\lambda^{-1}}.
\end{array}
$$
So, in this case $\,\lambda\log I(\lambda)$ does not vanish as $\lambda\rightarrow+0$ and the left inequality in (\ref{int-est}) shows that (\ref{last-eq}) is not valid. \rule{5pt}{5pt}

\medskip

{\bf Acknowledgments.}  I am grateful to A.S.Holevo and G.G.Amosov for useful remarks. I am also grateful to the  participants of the workshop "Recent advances in continuous variable quantum information theory", Barcelona, April, 2016  (especially to A.Winter) for stimulating discussion. Special thanks to the unknown referee for the valuable suggestions.

\end{document}